\documentclass[%
 reprint,
nofootinbib,
 amsmath,amssymb,
 aps,
 pre
]{revtex4-2}

\usepackage{graphicx}
\graphicspath{{figures/}} 
\usepackage{dcolumn}
\usepackage{bm}
\usepackage[dvipsnames]{xcolor}

\renewcommand{\Re}{\mathrm{Re}}
\newcommand{\Rm}{\mathrm{Rm}}
\newcommand{\kxmin}{k_{x, \mathrm{min}}}

\newcommand{\kxopt}{k_{x, \mathrm{opt}}}
\newcommand{\kyopt}{k_{y, \mathrm{opt}}}
\newcommand{\Gopt}{G_{\mathrm{opt}}}

\newcommand{\MA}{\mathrm{M_A}}

\begin{document}

\preprint{APS/123-QED}

\title{Nonmodal growth and optimal perturbations in magnetohydrodynamic shear flows}

\author{Adrian E.~Fraser}
\email{adrian.fraser@colorado.edu}
\affiliation{Department of Applied Mathematics, University of Colorado, Boulder, CO, USA}
\affiliation{Department of Astrophysical and Planetary Sciences, University of Colorado, Boulder, CO, USA}
\affiliation{Laboratory for Atmospheric and Space Physics, University of Colorado, Boulder, CO, USA}

\author{Alexis K. Kaminski}
\affiliation{
Department of Mechanical Engineering, University of California, Berkeley, Berkeley, CA, USA
}%
\author{Jeffrey S. Oishi}
\affiliation{
Department of Mechanical Engineering, University of New Hampshire, Durham, NH, USA
}%

\date{\today}

\begin{abstract}
In astrophysical shear flows, the Kelvin-Helmholtz (KH) instability is generally suppressed by magnetic tension provided a sufficiently strong streamwise magnetic field. This is often used to infer upper (or lower) bounds on field strengths in systems where shear-driven fluctuations are (or are not) observed, on the basis that perturbations cannot grow in the absence of linear instability. On the contrary, by calculating the maximum growth that small-amplitude perturbations can achieve in finite time for such a system, we show that perturbations can grow in energy by orders of magnitude even when the flow is sub-Alfv\'enic, raising the possibility that shear-driven turbulence may be found even in the presence of strong magnetic fields, and challenging inferences from the observed presence or absence of shear-driven fluctuations. We further show that magnetic fields introduce additional nonmodal growth mechanisms relative to the hydrodynamic case, and that 2D simulations miss key aspects of these growth mechanisms.
\end{abstract}

\maketitle


Shear flows are ubiquitous in astrophysical \cite{BNS_KH_nature,BNS_KH_PRD,rieger_KH_jet_review,Spiegel_Zahn_Tachocline,Balbus_Hawley}, space \cite{faganello_magnetized_2017,Hillier}, and fusion \cite{Terry_ZF} plasmas, where they drive or fundamentally alter turbulent fluctuations. These fluctuations provide a source of mixing in stellar interiors \cite{Garaud_LowPr_review}, 
may provide a source of coronal/chromospheric heating \cite{Hillier}, 
are associated with transport barriers in high-confinement regimes of tokamak plasmas \cite{burrell_effects_1997}, and constrain the efficiency of inertial confinement fusion designs \cite{zhou_instabilities_2024,zhou_hydrodynamic_2024,zhou_rayleightaylor_2017}. 

Often, the impact of flow shear on such fluctuations is viewed through the lens of normal mode stability analyses. 
A variety of shear-driven modal instabilities exist \cite{Velikhov,Chandrasekhar_MRI,Acheson,Balbus_Hawley,Goldreich_Schubert,Fricke,barker_angular_2019}, 
most famously the Kelvin-Helmholtz (KH) instability \cite{chandrasekhar,drazin}. 
Results from modal stability calculations---including parameters for which a system is unstable, and the length- and time-scales of the instability---are believed to give insight into the fully nonlinear evolution of the system. If a given flow is 
modally stable, 
then fluctuations and turbulence 
\textcolor{black}{have traditionally been} expected to be suppressed as well. 
In astrophysical shear layers, where KH is generally stable for sub-Alfv\'enic flows (i.e., where the Alfv\'en speed exceeds the change in flow velocity across the layer) \cite{chandrasekhar,Miura_Pritchett}, such arguments are routinely used to suggest that an absence of shear-driven 
fluctuations or KH billows implies a lower bound on the magnetic field strength 
\cite{Perseus_KH,cai_blazar_KH,fan_blazars_KH,chow_kelvinhelmholtz_2023}, based on the assumption that no \textcolor{black}{linear perturbations} can grow if the field is strong enough to make the flow sub-Alfv\'enic. Similarly, Squire's theorem that spanwise-independent modes are the most unstable
is often used (sometimes mistakenly \cite{hunt_1966,hughes_tobias}) to argue that 2D nonlinear simulations are sufficient to characterize fluctuations in such plasmas. 

However, this line of reasoning can be misleading. In the fluids literature, numerous studies have shown that \textcolor{black}{perturbations} can grow and drive turbulence and transport even in systems that are stable 
to modal instabilities. This phenomenon is broadly referred to as nonmodal or non-normal growth \cite{schmid_review}. 
In shear flows, two distinct mechanisms are known to drive nonmodal growth: the Orr mechanism \cite{Orr1907a,Farrell1988,farrell_1993}, which rotates and amplifies eddies tilted against the shear, and the lift-up effect \cite{Ellingsen1975,brandt2014}, which amplifies \textcolor{black}{perturbations} via advection of horizontal momentum. These mechanisms arise in all manner of shear flows---including shear layers, wall-bounded flows, and jets \cite{Butler1992,Trefethen1993,arratia,Arratia2013b,pickering_lift-up_2020}---where they 
\textcolor{black}{amplify perturbations} that differ significantly from those driven by modal growth mechanisms. For instance, while the fastest-growing KH modes in 
parallel shear flows are spanwise-invariant, that need not be the case for nonmodal growth: 3D \textcolor{black}{perturbations} can be significant and may even attain the most growth over intermediate times \cite{arratia}. Indeed, while the Orr mechanism can drive growth in both 2D (spanwise-invariant) and 3D \textcolor{black}{perturbations}, the lift-up effect drives 
\textcolor{black}{perturbations} with no \emph{streamwise} variation! 
Furthermore, nonmodal growth can 
\textcolor{black}{amplify perturbations} in flows that are linearly stable by classical modal stability analyses. 
For instance, sufficiently strong vertical density stratification 
stabilizes KH 
in neutral fluids (where buoyancy can stabilize KH just as magnetic tension can in MHD); however, in such flows, nonmodal mechanisms can lead to large 
\textcolor{black}{growth} of fluctuations \cite{Farrell1993b,kaminski_2014} which can significantly modify the background state \cite{kaminski_2017}---even for flows far from the stability boundary. This raises the question of whether 
sub-Alfv\'enic flows in astrophysical systems 
can support shear-driven fluctuations despite strong magnetic tension, just as stratified flows 
support such fluctuations despite strong stratification.


While nonmodal growth has been investigated in 
several plasma systems---including MRI \cite{squire_PRL,squire_ApJ}, tearing instability \cite{mactaggart}, and 
plasma drift waves \cite{friedman_PRL,landreman,friedman_HW}---the ubiquitous nonmodal effects seen in neutral-fluid shear layers have not, to our knowledge, been investigated in astrophysically relevant plasma models, despite their significance in the former. 
Implications for shear flows in liquid-metal duct flows have been widely explored\textcolor{black}{, often using methods that extend beyond the linear dynamics we consider here} \cite{Ref3_1,Ref3_2,Ref3_3,Ref3_4,Ref3_5,Ref3_6,Ref3_7,Ref3_8,Ref3_9}. However, work in that area 
\textcolor{black}{typically} considers 
wall-normal magnetic fields (not streamwise, as in the astrophysically relevant case 
considered here), 
\textcolor{black}{often in the limit of} 
large resistivity. That limit is largely irrelevant to astrophysical plasmas---which are nearly ideal conductors---and\textcolor{black}{, in this system, would preclude} two 
features 
we show to be quite important: Alfv\'en waves, and the transfer of energy from magnetic to kinetic. \footnote{\textcolor{black}{Note that some counterexamples exist in other systems where magnetic waves persist in the limit of large resistivity, see, e.g., Refs.~\cite{horn_2022,lalloz_alfven_2025,Skoutnev}; we note that those studies consider systems with additional physics such as rotation or injected current density at the wall, which are not included in the system considered here.}}

Thus, we investigate nonmodal growth in a shear layer with a streamwise magnetic field and finite resistivity. 
This system has multiple qualitative similarities with stratified shear layers, making it 
instructive to compare the two. Relative to the simplest unstratified shear layers in neutral fluids (where the only force that can stabilize an otherwise-unstable layer is viscosity), stratification and magnetic fields each give rise to a new force in the momentum equation that provides waves (internal gravity waves and Alfv\'en waves, respectively) and, in some limits, suppresses KH. 
We demonstrate that, as with the stratified 
case, significant perturbation growth can occur even when KH is suppressed by magnetic tension. We also encounter new nonmodal growth mechanisms introduced by the magnetic field.

\emph{Methods.---}We consider the growth of small-amplitude 3D disturbances $\mathbf{u}$, $\mathbf{b}$ to the background shear flow $\mathbf{U}_0 = U_0 \tanh(z/d) \hat{\mathbf{e}}_x$ and uniform, streamwise magnetic field $\mathbf{B}_0 = B_0 \hat{\mathbf{e}}_x$ in the framework of incompressible MHD with finite viscosity and resistivity. 
The equations governing these disturbances, linearized about $\mathbf{U}_0$ and $\mathbf{B}_0$ and non-dimensionalized using the 
layer half-width $d$, flow speed $U_0$, and 
field strength $B_0$, are
\begin{equation} \label{eq:lin-momentum}
\frac{\partial}{\partial t} \mathbf{u} + wU' \hat{\mathbf{e}}_x + U \frac{\partial}{\partial x}\mathbf{u} = - \nabla p + \frac{1}{\MA^2} \frac{\partial}{\partial x} \mathbf{b} + \frac{1}{\Re} \nabla^2 \mathbf{u},
\end{equation}
\begin{equation} \label{eq:lin-induction}
\frac{\partial}{\partial t} \mathbf{b} + U \frac{\partial}{\partial x}\mathbf{b} = \frac{\partial}{\partial x} \mathbf{u} + b_z U' \hat{\mathbf{e}}_x + \frac{1}{\Rm} \nabla^2 \mathbf{b},
\end{equation}
with $\nabla \cdot \mathbf{u} = 0$ and $\nabla \cdot \mathbf{b} = 0$.
Here, $\MA \equiv U_0/v_A$ is the Alfv\'{e}n Mach number with 
Alfv\'{e}n speed $v_A \propto B_0$, 
$\Re \equiv U_0 d / \nu$ 
the Reynolds number with 
kinematic viscosity $\nu$, $\Rm \equiv U_0 d / \eta$ 
the magnetic Reynolds number with 
magnetic diffusivity $\eta$, and $U' \equiv dU/dz$. 
Throughout this work, we 
take $\Re = \Rm = 250$. 
We choose these values because they are the largest we can achieve before numerical costs force us to reduce the breadth of our scans across $k_x$, $k_y$, and $M_A$, and fix $\Re/\Rm = 1$ because of its ubiquity in astrophysical studies (despite not being 
ubiquitous in astrophysics \cite{rincon_dynamo_2019}). We note that nonmodal growth in the hydrodynamic case 
depends on $\Re$ \cite{Butler1992}, but leave the question of how our results vary with $\Re$ and $\Rm$ to future work.
We impose periodic boundary conditions in $x$ and $y$, and no-slip, perfectly conducting boundaries in $z$. 
To enforce $\nabla \cdot \mathbf{b} = 0$, we use the 
vector potential in the Coulomb gauge (see, e.g., Ref.~\cite{Cresswell}). 

In contrast to modal stability analyses, we frame our stability problem by seeking an initial condition that maximizes some measure of perturbation growth---the ``linear optimal perturbation" (LOP). Let $\mathbf{X}(t) = [\mathbf{u}(t), \mathbf{b}(t)]^T$ represent the system state at time $t$. 
We define the ``gain'' $G_\chi (t_0)$ as the maximum amplification of some norm $|| \cdot ||_\chi^2$ that 
any initial condition can achieve by the target time $t_0$, i.e.,
\begin{equation}
    G_\chi(t_0) = \max_{\mathbf{X}(0) \neq 0} \frac{|| \mathbf{X}(t_0) ||_\chi^2}{||\mathbf{X}(0)||_\chi^2}.
\end{equation}
This gives our measure of perturbation growth to be maximized between $t = 0$ and $t_0$. The initial perturbation $\mathbf{X}(0)$ that achieves this amplification is the LOP, and $\mathbf{X}(t_0)$ is the ``evolved state.''

We measure growth using the energy norm
\begin{equation}
    E = \int d^3x \left[ |\mathbf{u}|^2 + \frac{1}{\MA^2} |\mathbf{B}|^2 \right],
\end{equation}
where the first term is twice the kinetic energy $K$ and the second term 
twice the magnetic energy $M$. 
In the absence of 
viscosity and resistivity, 
\begin{equation}\label{eq:energy-production}
    \frac{\partial}{\partial t} E = 2\int d^3x \underbrace{[- uwU'}_\text{KSP} + \underbrace{\MA^{-2} b_x b_z U']}_\text{MSP}\, .
\end{equation}
The first term on the right describes changes in kinetic energy via the background shear, i.e., ``kinetic shear production'' (KSP). The second term corresponds to growth via interactions between the background shear and 
magnetic field \textcolor{black}{perturbations}, 
which we call ``magnetic shear production'' (MSP). This term presents a significant departure from stratified neutral fluids 
and from the $\Rm \to 0$ 
limit, where only the KSP term is present, and can give rise to significant nonmodal growth in the system considered here. 

We calculate the LOPs and gains numerically following methods employed by Refs.~\cite{reddy_1993,squire_ApJ,mactaggart} and described in greater detail in the 
\textcolor{black}{Appendix}. In short, we construct the propagator 
using the eigenmodes of Eqs.~\eqref{eq:lin-momentum}--\eqref{eq:lin-induction}; the gain is then given by the largest singular value of the propagator. We expand our system in Fourier modes in $x$ and $y$ with horizontal wavenumber $\mathbf{k} = (k_x, k_y)$. Because 
the dynamics at each $\mathbf{k}$ are entirely decoupled,
the LOP and gain can be defined separately for each $\mathbf{k}$. For every $t_0$, we let $(\kxopt, \kyopt)$ denote the wavenumbers 
that maximize $G(t_0, k_x, k_y)$, and define $\Gopt(t_0) \equiv G(t_0, \kxopt, \kyopt)$.

Before proceeding to our results, key differences relative to stratified shear flows can already be anticipated. First, as noted above, this system includes a second energy production term in Eq.~\eqref{eq:energy-production} beyond the standard KSP term. Additionally, for perturbations invariant along $\mathbf{B}_0$ (so $k_x = 0$, or $\mathbf{k} \cdot \mathbf{B}_0 = 0$ 
more generally), Eqs.~\eqref{eq:lin-momentum} and \eqref{eq:lin-induction} decouple. 
Thus, for these perturbations, the evolution of $\mathbf{u}$ is identical to the hydrodynamic case regardless of 
field strength. 
An immediate consequence is that (provided viscosity is 
sufficiently small \cite{farrell_1993,brandt2014}) \emph{in the presence of uniform, horizontal magnetic fields, incompressible shear flows always permit some nonmodal growth regardless of field strength.} 
In our particular case, we thus expect the standard lift-up effect \cite{Ellingsen1975} at $k_x=0$ for all $\MA$. 

To investigate the role of domain size, 
we compare two sets of calculations. In our ``small box'' cases, we consider evenly spaced $k_x\in[0.1,1]$ and $k_y\in[0,1]$. The minimum $k_x$ in these cases is comparable to the wavenumber of the most unstable 
KH mode for weaker magnetic fields, and roughly mimics the smallest nonzero $k_x$ found in many direct numerical simulations (DNS) of this system (e.g., Ref.~\cite{fraser_impact_2021}). Thus, these calculations are aimed at identifying nonmodal growth mechanisms one might anticipate at nonzero $k_x$ in standard DNS calculations. In our ``large box'' cases, we consider longer-wavelength modes with logarithmically spaced $k_x \in [10^{-3}, 1]$ 
(and evenly spaced $k_y\in[0,0.775]$). We find that, until $t_0$ becomes very large, $G$ is independent of $k_x$ for $k_x \lesssim 10^{-2}$. This lends us confidence that 
our results are not restricted by this floor on $k_x$, thus providing some insight into the $k_x=0$ case, which we are unable to calculate directly 
(see End Matter).

\begin{figure*}
\centering
\includegraphics[width=\textwidth]{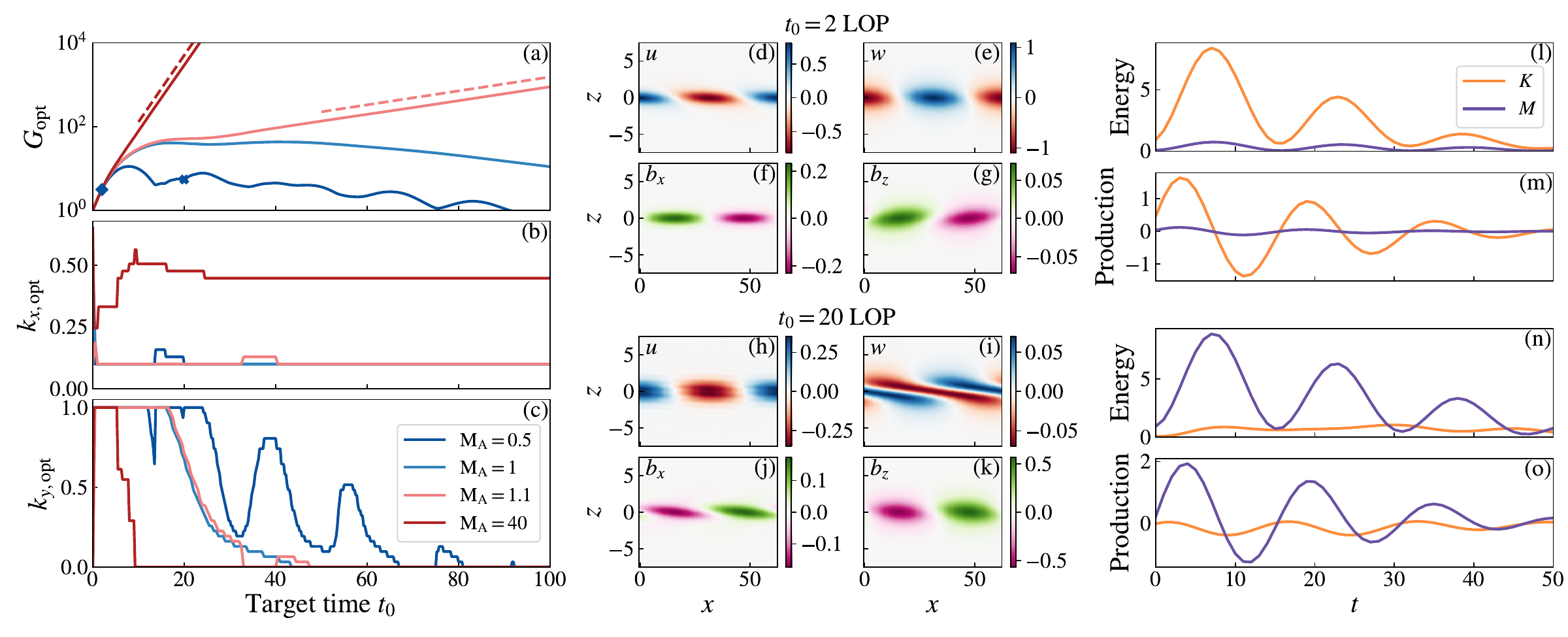}
\caption{Linear optimal perturbations, i.e., perturbations attaining the largest growth over a finite time, for the ``small box'' domain. (Left) Details of the LOPs for several values of $\MA$: (a) gains, (b) streamwise wavenumbers, and (c) spanwise wavenumbers. 
The points in panel (a) correspond to the cases shown in the middle column. (Middle) Spatial structure of $u$, $w$, $b_x$, and $b_z$ of the LOPs in the $xz$-plane. Panels (d)--(g) correspond to LOPs for a short target time $t_0=2$, while (h)--(k) correspond to an intermediate target time $t_0=20$. $\MA=0.5$ for both. (Right) Evolution of (l) kinetic and (n) magnetic energy components, and corresponding production terms (m), (o), for the LOPs highlighted in the middle column.}
\label{fig:smallbox_megaplot}
\end{figure*}


\emph{Small boxes.---}We first consider ``small-box'' LOPs, 
Fig.~\ref{fig:smallbox_megaplot}. 
Panel (a) shows optimal energy gains $\Gopt$ as a function of target time $t_0$. For 
$\MA>1$ (where normal-mode instability 
is possible), the gain curves 
follow the modal growth rate at large $t_0$ as expected \cite{schmid_review}, while substantial nonmodal growth is possible at 
smaller $t_0$ and for stronger magnetic fields (lower $\MA$). 
The peak gain for $\MA = 1$ approaches 43, comparable to the marginally-stable case in Ref.~\cite{kaminski_2014} (which computed LOPs for a uniformly-stratified shear layer where, as in the case considered here, the addition of a new force can stabilize KH). 
Just as stratification affected peak gain in that study, we find stronger magnetic fields reduce maximum growth here. 

The preferred structures also share similarities with the stratified case. The wavenumbers $\mathbf{k}_\mathrm{opt}$ corresponding to $\Gopt(t_0)$ are given in Fig.~\ref{fig:smallbox_megaplot}(b-c), and the spatial structures in the $xz$-plane for two different LOPs are shown in Fig.~\ref{fig:smallbox_megaplot}(d-k). At late times, the spanwise wavenumber $k_{y,\mathrm{opt}}\rightarrow 0$, consistent with the 
most-unstable normal modes which are 2D in the $xz$-plane. Meanwhile, the large gains associated with shorter $t_0$ are inherently 3D, with large $k_y$ comparable to those seen in neutral fluids \cite{arratia,kaminski_2014} \textcolor{black}{(note that the observed plateau where $\kyopt \approx 1$ at short $t_0$ is a consequence of taking $k_{y, \mathrm{max}} = 1$; when larger $k_{y, \mathrm{max}}$ were considered, we found $\Gopt$ changed by less than 10\% even when $\kyopt$ was almost three times larger)}. 
The vertical structures of the LOPs resemble eddies tilted against the background shear, allowing for transient growth via the Orr mechanism \cite{Orr1907a,farrell_1993,Tearle2004}. 

We also note key differences relative to the stratified case. First, MSP 
drives growth for some LOPs. This is seen in panels (l-o), which show the time evolution of $K$ vs $M$ and of KSP vs MSP for the two LOPs shown in panels~(d-k). 
The $t_0=2$ LOP is largely driven by 
KSP and correspondingly has much more kinetic than magnetic energy; the converse is true for the $t_0=20$ LOP.
This is also reflected in the relative phases of the perturbation components: KSP is maximized for perturbations where $u$ and $w$ are \emph{anti-aligned} ($u \sim -w$), while MSP is maximized when $b_x$ and $b_z$ are \emph{aligned} 
[note 
each terms' sign in Eq.~\eqref{eq:energy-production}]. In general, 
KSP is preferred for short-$t_0$ perturbations, while 
for 
KH-stable flows at larger $t_0$, there is an oscillation between the two at about twice the Alfv\'en frequency. Thus, unlike the stratified case, this system permits two sources of growth that compete to determine the nature of the LOP for a given $t_0$. 
For either, perturbations eventually reach a state dominated by oscillations between $K$ and $M$ at near-equipartition, a characteristic feature of shear Alfv\'en waves.


Panel (b) reveals a final noteworthy 
feature of this system: for all KH-stable ($\MA \leq 1$) cases, $\kxopt \to \kxmin$ 
for all but the shortest $t_0$.
When the magnetic field is strong enough to suppress 
KH, not only are the LOPs \emph{not} 2D in the $xz$-plane, they \emph{also} become very extended in the streamwise direction. 
(This was not seen in the stratified 
case \cite{kaminski_2014}, where 
LOPs had a similar streamwise length scale 
to modally-unstable cases). While $\kxmin = 0.1$ corresponds to a domain length 
that is already larger than 
typically used in nonlinear simulations of this system, in what follows we extend to even smaller $\kxmin$ to identify how much growth is possible for larger-scale perturbations.

\begin{figure*}
\centering
\includegraphics[width=\textwidth]{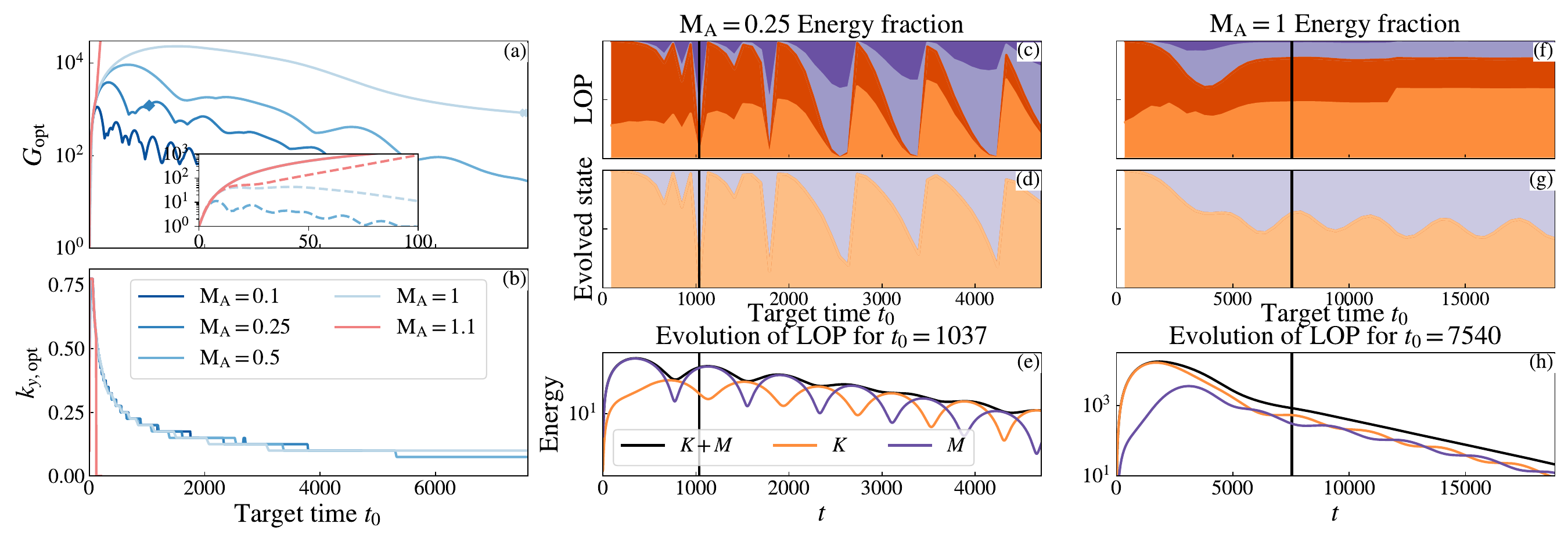}
\caption{Linear optimal perturbations for the ``large box'' domain. (a) LOP gains for different $\MA$ and a range of target times $t_0$; in the inset, large box calculations (solid) are compared against small box (dashed)---note all large-box curves overlap within this range. 
(b) Spanwise wavenumbers $\kyopt$ for the LOPs.
(c) Fraction of energy in different components for $\MA=0.25$ LOPs for different $t_0$.
Initially, energy is mostly in $v$ (orange), $w$ (dark orange), $b_y$ (purple), and $b_z$ (dark purple). (d) As for (c), but for the evolved state. The energy is now mostly in $u$ (light orange) and $b_x$ (light purple). (e) Evolution of kinetic (orange), magnetic (purple), and total (black) energy for the $t_0=1000$ LOP (vertical black lines in (c-e), blue diamond in (a)). 
(f-h) As in panels (c-e), but for $\MA=1$ and $t_0=7500$. 
}
\label{fig:bigbox_megaplot}
\end{figure*}

\emph{Large boxes.---}Key ``large box'' results are highlighted in Fig.~\ref{fig:bigbox_megaplot}. Several noteworthy differences from the ``small box'' calculations are immediately apparent. Linearly stable configurations ($\MA \leq 1$) can achieve orders of magnitude higher energy gains in these larger domains, 
with the $\MA = 1$ case exceeding $\Gopt = 10^4$, and even the strongly stabilized $\MA=0.1$ case achieving $\Gopt \approx 10^3$ (panel~a). 
Additionally, while $\kxopt \to 0$ (not shown) for $\MA \leq 1$ as before, $\kyopt$ remains finite.
This is the \emph{opposite} of 
expectations from modal stability analyses for shear flows with streamwise magnetic fields, where fastest-growing modes often have $k_y=0$ (e.g., Ref.~\cite{fraser_JFM_2022}).

That large-box MHD calculations achieve much higher gains than 
small boxes or the stratified case can be understood by 
recalling that the Lorentz force (and thus the stabilizing 
magnetic tension) vanishes from Eq.~\eqref{eq:lin-momentum} for modes with $\mathbf{k} \cdot \mathbf{B}_0 = 0$. 
One can therefore show 
that, in the inviscid limit, the streamwise velocity grows as
\begin{equation}
    u(t) = u(0) - w U' t,
\end{equation}
a phenomenon known as the lift-up effect \cite{Ellingsen1975,brandt2014}.
This effect provides a growth mechanism for $k_x = 0$ perturbations that is unaffected by the background field \emph{no matter its strength}. 
Thus, while the Lorentz and buoyancy forces both tend to stabilize streamwise KH modes, only the latter stabilizes KH \emph{and} the spanwise modes driven by lift-up.

Not only does $\mathbf{B}_0$ have no effect on hydrodynamic lift-up in this system for $k_x = 0$, but it introduces an additional growth mechanism. The MSP term in Eq.~\eqref{eq:energy-production} gives rise to 
solutions (in ideal MHD for $k_x=0$) that evolve as
\begin{equation}
    b_x(t) = b_x(0) + b_z U' t.
\end{equation}
Thus, shear 
also drives growth of \emph{magnetic} \textcolor{black}{perturbations}---a magnetic analogue of the hydrodynamic lift-up effect. Note that this is precisely the ``omega effect" relevant to dynamos realized in a Cartesian domain \cite{rincon_dynamo_2019,tobias_turbulent_2021}. 

The impact of these two lift-up effects is demonstrated by Fig.~\ref{fig:bigbox_megaplot}(c-d) and (f-g), which show that the energy in the LOPs consists of a combination of $w$ and $b_z$ (with some $v$ and $b_y$ contributions to satisfy $\nabla\cdot\mathbf{u}=0$ and $\nabla\cdot\mathbf{b}=0$), while the evolved states consist almost entirely of energy in $u$ and $b_x$. 
For short $t_0$, hydrodynamic lift-up is the dominant contributor. 
For larger $t_0$, either KH becomes the dominant growth mechanism (for $\MA > 1$, not shown) and the energy breakdown merely reflects that of the unstable mode, or the optimal growth arises from a combination of the two lift-up mechanisms. 
For $\MA = 1$, hydrodynamic lift-up remains preferred, 
but for $\MA < 1$ the balance between the two varies with $t_0$, 
see panels (c)--(d). 

We find that for each of the KH-stable ($\MA \leq 1$) cases, the global maximum in $\Gopt$ occurs at $t_0 = T_A/4$ (with Alfv\'en crossing time $T_A = 2\pi/\kxmin v_A$) suggesting that, at this $\Re$, the initial departure from the expected secular growth is due to the excitation of Alfv\'{e}n waves by the Lorentz force rather than dissipative effects as seen in the hydrodynamic case \cite{brandt2014,farrell_1993}. The transition from lift-up to waves is further demonstrated in Fig.~\ref{fig:bigbox_megaplot}(e) and (h), which show the $K$ (orange) and $M$ (purple) evolution in initial value calculations corresponding to the LOPs denoted by the symbols in panel (a). After the initial peak from lift-up, energy is continually exchanged between $K$ and $M$ at twice the Alfv\'{e}n frequency, consistent with the presence of a standing Alfv\'{e}n wave, as in the small-box case. 




\emph{Discussion.---}We have calculated the maximum energy growth attainable by small-amplitude perturbations to a shear layer with a streamwise, uniform magnetic field. While the dynamics as $t \to \infty$ are dictated by modal stability analyses, significant nonmodal growth is possible over finite time horizons. 
This is true even when the 
field is strong enough to suppress KH, 
with energy growing by a factor of over $10^4$ in some cases. Thus, when magnetized shear flows are observed in the laboratory or in nature, a strong, flow-aligned magnetic field need not imply that the configuration is stable and that no shear-driven 
fluctuations will grow\textcolor{black}{---particularly if, as has been demonstrated in the stratified case \cite{kaminski_2017}, this linear nonmodal growth is sufficient to drive large-amplitude fluctuations and nonlinear mixing}. This demonstrates the importance of advancing beyond modal stability analyses, the current state of the art in understanding shear-driven fluctuations in astrophysical flows.

Furthermore, while spanwise-uniform ($k_y=0$) modes 
are often 
the most unstable according to traditional modal stability analyses---an observation often used to motivate 2D simulations of plasma shear flows---our results show that the perturbations that grow the most may be either 3D or (nearly) uniform in the \emph{streamwise} direction (with $k_x \approx 0$). This finding suggests that simulations that neglect spanwise variation based on the results of modal stability analyses may miss key sources of \textcolor{black}{perturbation} growth, and thus we expect more \textcolor{black}{perturbation} growth in 3D simulations than in 2D, particularly for $M_A \lesssim 1$.

\textbf{Data availability.---} Software used to create datasets and figures is hosted at Zenodo \cite{fraser_linopt_code} and can be found at \href{https://github.com/afraser3/nonmodal-MHD-KH}{this} GitHub repository.

We acknowledge helpful discussions with I.~Grooms, B.~Hindman, C.~Caulfield, and B.~Tripathi. 
AEF was supported by NSF awards AST-1814327 and AST-1908338, NASA HTMS grant 80NSSC20K1280, and the George Ellery Hale Postdoctoral Fellowship in Solar, Stellar and Space Physics at the University of Colorado, Boulder. 
AKK was supported as the Ho-Shang and Mei-Li Lee Faculty Fellow at UC Berkeley.
JSO was supported by NASA HTMS grant 80NSSC20K1280 and NASA OSTFL Grant 80NSSC22K1738.
This work grew out of discussions during the 2021 Kavli Institute for Theoretical Physics workshop on ``Layering in Atmospheres, Oceans and Plasmas'' (supported by NSF grant PHY-2309135). 
Computations were conducted with support of the NASA High End Computing (HEC) Program through the NASA Advanced Supercomputing (NAS) Division at Ames Research Center on Pleiades, the Lux supercomputer at UC Santa Cruz, funded by NSF MRI grant AST-1828315, and the Alpine cluster, jointly funded by the University of Colorado Boulder, the University of Colorado Anschutz, Colorado State University, and the National Science Foundation (award 2201538).

\appendix
\section*{Appendix}

In contrast to modal stability analyses, here we frame our stability problem by seeking an initial condition that maximizes some measure of perturbation growth---the ``linear optimal perturbation'' (LOP). 
Consider the system
\begin{equation}
\frac{\partial \mathbf{X}}{\partial t} = \mathcal{L} \mathbf{X},
\label{eq:simplest_system}
\end{equation}
where $\mathbf{X}(t)$ is the system state and $\mathcal{L}$ is a time-independent linear differential operator.
Defining the propagator $\mathcal{K}(t) = \exp(\mathcal{L} t)$, the general solution at some time $t$ can be written 
\begin{equation}
\mathbf{X}(t) = \mathcal{K}(t) \mathbf{X}(0)\, .
\label{eq:simplest_solution}
\end{equation}
The maximum amplification in terms of some norm $|| \cdot ||_\chi^2$ by the target time $t_0$, referred to as the ``gain", is then
\begin{equation}
G_\chi(t_0) = \max_{\mathbf{X}(0) \neq 0} \frac{|| \mathcal{K}(t_0) \mathbf{X}(0) ||_\chi^2}{||\mathbf{X}(0)||_\chi^2}.
\end{equation}
This gives our measure of perturbation growth between $t=0$ and some nonzero target time $t_0$ that we seek to maximize. 

We perform our calculations numerically using the pseudospectral tau method as implemented in the Dedalus framework \cite{Dedalus_methods} leveraging its eigentools package \cite{Eigentools}. 
We expand in Fourier modes in $x$ and $y$ with horizontal wavenumber $\mathbf{k} = (k_x, k_y)$ 
and expand in Chebyshev polynomials in $z$.
From here, our calculation largely follows Ref.~\cite{squire_ApJ}: we calculate $\mathcal{K}$ at each $\mathbf{k}$ in the eigenmode basis, 
and calculate the gain and LOP using a singular value decomposition, specifying the energy norm $E$ using a Cholesky decomposition. We identify spurious eigenmodes arising from discretization using the spurious mode rejection algorithm \cite{boyd_chebyshev_2001} implemented in eigentools, and remove them from the calculation of gains and LOPs using the method described in Ref.~\cite{mactaggart}.
To expedite calculations, we use sparse linear algebra solvers to calculate only a subset of the eigenmodes of the system. Rigorous convergence checks have been performed to ensure our results do not change significantly upon increased resolution in $z$ or increased number of eigenmodes obtained by the solver. We find that calculations with $N_z = 512$ Chebyshev modes and 600 eigenmodes are well converged for all parameters reported here (where 600 corresponds to total number of eigenmodes obtained by the routine before spurious modes are identified and removed). While this method provides robust results for all parameters shown in this paper, we find that the condition number of $\mathcal{L}$ grows significantly as $k_x \to 0$, and thus we are unable to reliably calculate gains or LOPs for $k_x \lesssim 10^{-4}$ using double-precision arithmetic (note Ref.~\cite{squire_ApJ} used extended precision).

\bibliography{mhd_linopt}

@ARTICLE{Eigentools,
       author = {{Oishi}, Jeffrey and {Burns}, Keaton and {Clark}, S. and {Anders}, Evan and {Brown}, Benjamin and {Vasil}, Geoffrey and {Lecoanet}, Daniel},
        title = "{eigentools: A Python package for studying differential eigenvalue problems with an emphasis on robustness}",
      journal = {The Journal of Open Source Software},
         year = 2021,
        month = jun,
       volume = {6},
       number = {62},
          eid = {3079},
        pages = {3079},
          doi = {10.21105/joss.03079},
       adsurl = {https://ui.adsabs.harvard.edu/abs/2021JOSS....6.3079O}
}

@ARTICLE{Dedalus_methods,
       author = {{Burns}, Keaton J. and {Vasil}, Geoffrey M. and {Oishi}, Jeffrey S. and {Lecoanet}, Daniel and {Brown}, Benjamin P.},
        title = "{Dedalus: A flexible framework for numerical simulations with spectral methods}",
      journal = {Physical Review Research},
         year = 2020,
        month = apr,
       volume = {2},
       number = {2},
          eid = {023068},
        pages = {023068},
          doi = {10.1103/PhysRevResearch.2.023068},
archivePrefix = {arXiv},
       eprint = {1905.10388},
 primaryClass = {astro-ph.IM},
       adsurl = {https://ui.adsabs.harvard.edu/abs/2020PhRvR...2b3068B}
}

@article{fraser_JFM_2022,
	title = {Non-ideal instabilities in sinusoidal shear flows with a streamwise magnetic field},
	volume = {949},
	issn = {0022-1120, 1469-7645},
	url = {https://www.cambridge.org/core/journals/journal-of-fluid-mechanics/article/nonideal-instabilities-in-sinusoidal-shear-flows-with-a-streamwise-magnetic-field/3A4915E1FF6A94E4FF86B7640FE77F26},
	doi = {10.1017/jfm.2022.782},
	urldate = {2022-10-06},
	journal = {Journal of Fluid Mechanics},
	author = {Fraser, A. E. and Cresswell, I. G. and Garaud, P.},
	month = oct,
	year = {2022},
	pages = {A43}
}

@article{Cresswell,
	title = {Force balances in strong-field magnetoconvection simulations},
	volume = {8},
	url = {https://link.aps.org/doi/10.1103/PhysRevFluids.8.093503},
	doi = {10.1103/PhysRevFluids.8.093503},
	number = {9},
	urldate = {2023-10-16},
	journal = {Physical Review Fluids},
	author = {Cresswell, Imogen G. and Anders, Evan H. and Brown, Benjamin P. and Oishi, Jeffrey S. and Vasil, Geoffrey M.},
	month = sep,
	year = {2023},
	pages = {093503}
}

@article{fraser_impact_2021,
	title = {The impact of magnetic fields on momentum transport and saturation of shear-flow instability by stable modes},
	volume = {28},
	issn = {1070-664X},
	url = {https://aip.scitation.org/doi/abs/10.1063/5.0034575},
	doi = {10.1063/5.0034575},
	number = {2},
	urldate = {2021-07-12},
	journal = {Physics of Plasmas},
	author = {Fraser, A. E. and Terry, P. W. and Zweibel, E. G. and Pueschel, M. J. and Schroeder, J. M.},
	month = feb,
	year = {2021},
	pages = {022309}
}

@article{kaminski_2014,
	title = {Transient growth in strongly stratified shear layers},
	volume = {758},
	issn = {0022-1120, 1469-7645},
	url = {https://www.cambridge.org/core/journals/journal-of-fluid-mechanics/article/transient-growth-in-strongly-stratified-shear-layers/E03CFAFDF54AFC26E5E9BCCA71FDB53F},
	doi = {10.1017/jfm.2014.552},
	urldate = {2021-10-29},
	journal = {Journal of Fluid Mechanics},
	author = {Kaminski, A. K. and Caulfield, C. P. and Taylor, J. R.},
	month = nov,
	year = {2014}
}

@article{kaminski_2017,
	title = {Nonlinear evolution of linear optimal perturbations of strongly stratified shear layers},
	volume = {825},
	issn = {0022-1120, 1469-7645},
	url = {https://www.cambridge.org/core/journals/journal-of-fluid-mechanics/article/nonlinear-evolution-of-linear-optimal-perturbations-of-strongly-stratified-shear-layers/33C775C051CB4736FB76B7A272C6A0C6},
	doi = {10.1017/jfm.2017.396},
	urldate = {2021-10-29},
	journal = {Journal of Fluid Mechanics},
	author = {Kaminski, A. K. and Caulfield, C. P. and Taylor, J. R.},
	month = aug,
	year = {2017},
	pages = {213--244}
}

@article{arratia,
	title = {Transient perturbation growth in time-dependent mixing layers},
	volume = {717},
	issn = {0022-1120, 1469-7645},
	url = {https://www.cambridge.org/core/journals/journal-of-fluid-mechanics/article/transient-perturbation-growth-in-timedependent-mixing-layers/3AF6992F0F57F8375FD82B314BEDCA13},
	doi = {10.1017/jfm.2012.562},
	urldate = {2022-10-07},
	journal = {Journal of Fluid Mechanics},
	author = {Arratia, C. and Caulfield, C. P. and Chomaz, J.-M.},
	month = feb,
	year = {2013},
	pages = {90--133}
}

@article{reddy_1993,
	title = {Pseudospectra of the {Orr}–{Sommerfeld} {Operator}},
	volume = {53},
	issn = {0036-1399},
	url = {https://epubs.siam.org/doi/10.1137/0153002},
	doi = {10.1137/0153002},
	number = {1},
	urldate = {2020-12-24},
	journal = {SIAM Journal on Applied Mathematics},
	author = {Reddy, Satish C. and Schmid, Peter J. and Henningson, Dan S.},
	month = feb,
	year = {1993},
	pages = {15--47}
}

@article{Orr1907a,
	author = {Orr, W. M'F.},
	title = {The stability or instability of the steady motions of a perfect liquid and of a viscous liquid. Part {I}: A perfect liquid.},
	journal = {Proc. Roy. Irish Acad. A},
	volume = {27},
	pages = {9-68},
	year = {1907}
}

@article{farrell_1993,
	title = {Optimal excitation of three‐dimensional perturbations in viscous constant shear flow},
	volume = {5},
	issn = {0899-8213},
	url = {https://doi.org/10.1063/1.858574},
	doi = {10.1063/1.858574},
	number = {6},
	urldate = {2024-02-15},
	journal = {Physics of Fluids A: Fluid Dynamics},
	author = {Farrell, Brian F. and Ioannou, Petros J.},
	month = jun,
	year = {1993},
	pages = {1390--1400}
}

@article{mactaggart,
	title = {The non-modal onset of the tearing instability},
	volume = {84},
	issn = {1469-7807},
	url = {http://dx.doi.org/10.1017/S0022377818001009},
	doi = {10.1017/s0022377818001009},
	number = {5},
	journal = {Journal of Plasma Physics},
	author = {MacTaggart, D.},
	month = sep,
	year = {2018},
}

@article{squire_ApJ,
	title = {Magnetorotational instability: nonmodal growth and the relationship of global modes to the shearing box},
	volume = {797},
	issn = {0004-637X},
	shorttitle = {{MAGNETOROTATIONAL} {INSTABILITY}},
	url = {https://doi.org/10.1088/0004-637x/797/1/67},
	doi = {10.1088/0004-637X/797/1/67},
	number = {1},
	urldate = {2021-09-21},
	journal = {The Astrophysical Journal},
	author = {Squire, J. and Bhattacharjee, A.},
	month = nov,
	year = {2014},
	pages = {67}
}

@article{squire_PRL,
	title = {Nonmodal {Growth} of the {Magnetorotational} {Instability}},
	volume = {113},
	url = {https://link.aps.org/doi/10.1103/PhysRevLett.113.025006},
	doi = {10.1103/PhysRevLett.113.025006},
	number = {2},
	urldate = {2021-09-21},
	journal = {Physical Review Letters},
	author = {Squire, J. and Bhattacharjee, A.},
	month = jul,
	year = {2014},
	pages = {025006}
}

@article{hunt_1966,
	title = {On the {Stability} of {Parallel} {Flows} with {Parallel} {Magnetic} {Fields}},
	volume = {293},
	issn = {0080-46301364-5021},
	url = {https://ui.adsabs.harvard.edu/abs/1966RSPSA.293..342H},
	doi = {10.1098/rspa.1966.0175},
	urldate = {2021-12-21},
	journal = {Proceedings of the Royal Society of London Series A},
	author = {Hunt, J. C. R.},
	month = aug,
	year = {1966},
	pages = {342--358}
}

@article{hughes_tobias,
	title = {On the instability of magnetohydrodynamic shear flows},
	volume = {457},
	issn = {1364-5021},
	url = {https://royalsocietypublishing.org/doi/10.1098/rspa.2000.0725},
	doi = {10.1098/rspa.2000.0725},
	number = {2010},
	journal = {Proceedings of the Royal Society of London. Series A: Mathematical, Physical and Engineering Sciences},
	author = {Hughes, D.W. and Tobias, S.M.},
	month = jun,
	year = {2001},
	pages = {1365--1384}
}

@book{boyd_chebyshev_2001,
	series = {Dover {Books} on {Mathematics}},
	title = {Chebyshev and {Fourier} {Spectral} {Methods}: {Second} {Revised} {Edition}},
	isbn = {978-0-486-41183-5},
	url = {https://books.google.com/books?id=lEWnQWyzLQYC},
	publisher = {Dover Publications},
	author = {Boyd, J.P.},
	year = {2001},
	lccn = {lc00027475},
}

@article{schmid_review,
	title = {Nonmodal {Stability} {Theory}},
	volume = {39},
	url = {https://doi.org/10.1146/annurev.fluid.38.050304.092139},
	doi = {10.1146/annurev.fluid.38.050304.092139},
	number = {1},
	urldate = {2020-12-24},
	journal = {Annual Review of Fluid Mechanics},
	author = {Schmid, Peter J.},
	year = {2007},
	pages = {129--162},
}

@article{Ellingsen1975,
	author = {Ellingsen, T. and Palm, E.},
	doi = {10.1063/1.861156},
	title = {Stability of linear flow},
	journal = {Phys. Fluids},
	volume = {18},
	number = {4},
	pages = {487-488},
	year = {1975}
}

@article{brandt2014,
	series = {Enok {Palm} {Memorial} {Volume}},
	title = {The lift-up effect: {The} linear mechanism behind transition and turbulence in shear flows},
	volume = {47},
	issn = {0997-7546},
	shorttitle = {The lift-up effect},
	url = {https://www.sciencedirect.com/science/article/pii/S0997754614000405},
	doi = {10.1016/j.euromechflu.2014.03.005},
	urldate = {2023-08-24},
	journal = {European Journal of Mechanics - B/Fluids},
	author = {Brandt, Luca},
	month = sep,
	year = {2014},
	pages = {80--96}
}

@article{rincon_dynamo_2019,
	title = {Dynamo theories},
	volume = {85},
	issn = {0022-3778, 1469-7807},
	url = {https://www.cambridge.org/core/journals/journal-of-plasma-physics/article/dynamo-theories/DEEA156E0E7DADE00DE61479D2DC2EED},
	doi = {10.1017/S0022377819000539},
	number = {4},
	journal = {Journal of Plasma Physics},
	author = {Rincon, François},
	month = aug,
	year = {2019},
	pages = {205850401}
}

@article{tobias_turbulent_2021,
	title = {The turbulent dynamo},
	volume = {912},
	issn = {0022-1120, 1469-7645},
	url = {https://www.cambridge.org/core/journals/journal-of-fluid-mechanics/article/turbulent-dynamo/8E03224E306A1DF360489DBA3DB09E06},
	doi = {10.1017/jfm.2020.1055},
	journal = {Journal of Fluid Mechanics},
	author = {Tobias, S. M.},
	month = apr,
	year = {2021},
	pages = {P1}
}

@ARTICLE{Velikhov,
       author = {{Velikhov}, Evgeny Pavlovich},
        title = "{Stability of an Ideally Conducting Liquid Flowing between Cylinders Rotating in a Magnetic Field}",
      journal = {Soviet Journal of Experimental and Theoretical Physics},
         year = 1959,
        month = nov,
       volume = {9},
       number = {5},
        pages = {995-998},
       adsurl = {https://ui.adsabs.harvard.edu/abs/1959JETP....9..995V}
}

@ARTICLE{Chandrasekhar_MRI,
       author = {{Chandrasekhar}, S.},
        title = "{The Stability of Non-Dissipative Couette Flow in Hydromagnetics}",
      journal = {Proceedings of the National Academy of Science},
         year = 1960,
        month = feb,
       volume = {46},
       number = {2},
        pages = {253-257},
          doi = {10.1073/pnas.46.2.253},
       adsurl = {https://ui.adsabs.harvard.edu/abs/1960PNAS...46..253C}
}

@ARTICLE{Acheson,
       author = {{Acheson}, D.~J. and {Gibbons}, M.~P.},
        title = "{On the Instability of Toroidal Magnetic Fields and Differential Rotation in Stars}",
      journal = {Philosophical Transactions of the Royal Society of London Series A},
         year = 1978,
        month = jun,
       volume = {289},
       number = {1363},
        pages = {459-500},
          doi = {10.1098/rsta.1978.0066},
       adsurl = {https://ui.adsabs.harvard.edu/abs/1978RSPTA.289..459A}
}

@ARTICLE{Balbus_Hawley,
       author = {{Balbus}, Steven A. and {Hawley}, John F.},
        title = "{A Powerful Local Shear Instability in Weakly Magnetized Disks. I. Linear Analysis}",
      journal = {The Astrophysical Journal},
         year = 1991,
        month = jul,
       volume = {376},
        pages = {214},
          doi = {10.1086/170270},
       adsurl = {https://ui.adsabs.harvard.edu/abs/1991ApJ...376..214B}
}

@ARTICLE{Goldreich_Schubert,
       author = {{Goldreich}, Peter and {Schubert}, Gerald},
        title = "{Differential Rotation in Stars}",
      journal = {The Astrophysical Journal},
         year = 1967,
        month = nov,
       volume = {150},
        pages = {571},
          doi = {10.1086/149360},
       adsurl = {https://ui.adsabs.harvard.edu/abs/1967ApJ...150..571G}
}

@ARTICLE{Fricke,
       author = {{Fricke}, K.},
        title = "{Instabilit{\"a}t station{\"a}rer Rotation in Sternen}",
      journal = {Zeitschrift für Astrophysik},
         year = 1968,
        month = jan,
       volume = {68},
        pages = {317},
       adsurl = {https://ui.adsabs.harvard.edu/abs/1968ZA.....68..317F}
}

@article{barker_angular_2019,
	title = {Angular momentum transport by the {GSF} instability: non-linear simulations at the equator},
	volume = {487},
	issn = {0035-8711},
	shorttitle = {Angular momentum transport by the {GSF} instability},
	url = {https://doi.org/10.1093/mnras/stz1386},
	doi = {10.1093/mnras/stz1386},
	number = {2},
	urldate = {2021-03-02},
	journal = {Monthly Notices of the Royal Astronomical Society},
	author = {Barker, A J and Jones, C A and Tobias, S M},
	month = aug,
	year = {2019},
	pages = {1777--1794}
}

@book{chandrasekhar,
	title = {Hydrodynamic and {Hydromagnetic} {Stability}},
	publisher = {Oxford University Press},
	author = {Chandrasekhar, S.},
	year = {1961},
}

@book{drazin,
	title = {Hydrodynamic {Stability}},
	publisher = {Cambridge University Press},
	author = {Drazin, P. G. and Reid, W.H.},
	year = {1981},
}

@ARTICLE{Miura_Pritchett,
       author = {{Miura}, A. and {Pritchett}, P.~L.},
        title = "{Nonlocal stability analysis of the MHD Kelvin-Helmholtz instability in a compressible plasma}",
      journal = {Journal of Geophysical Research},
         year = 1982,
        month = sep,
       volume = {87},
       number = {A9},
        pages = {7431-7444},
          doi = {10.1029/JA087iA09p07431},
       adsurl = {https://ui.adsabs.harvard.edu/abs/1982JGR....87.7431M}
}

@article{faganello_magnetized_2017,
	title = {Magnetized {Kelvin}–{Helmholtz} instability: theory and simulations in the {Earth}’s magnetosphere context},
	volume = {83},
	issn = {0022-3778, 1469-7807},
	shorttitle = {Magnetized {Kelvin}–{Helmholtz} instability},
	url = {https://www.cambridge.org/core/journals/journal-of-plasma-physics/article/magnetized-kelvinhelmholtz-instability-theory-and-simulations-in-the-earths-magnetosphere-context/3D1579EB6DDCF9B56A0FAD15AF8D74D1},
	doi = {10.1017/S0022377817000770},
	number = {6},
	urldate = {2023-08-24},
	journal = {Journal of Plasma Physics},
	author = {Faganello, Matteo and Califano, Francesco},
	month = dec,
	year = {2017},
	pages = {535830601}
}

@ARTICLE{Hillier,
       author = {{Hillier}, Andrew and {Arregui}, I{\~n}igo and {Matsumoto}, Takeshi},
        title = "{Nonlinear Wave Damping by Kelvin{\textendash}Helmholtz Instability-induced Turbulence}",
      journal = {The Astrophysical Journal},
         year = 2024,
        month = may,
       volume = {966},
       number = {1},
          eid = {68},
        pages = {68},
          doi = {10.3847/1538-4357/ad306f}
}

@article{landreman,
	title = {Generalized universal instability: transient linear amplification and subcritical turbulence},
	volume = {81},
	issn = {0022-3778, 1469-7807},
	shorttitle = {Generalized universal instability},
	url = {https://www.cambridge.org/core/journals/journal-of-plasma-physics/article/generalized-universal-instability-transient-linear-amplification-and-subcritical-turbulence/88A6FAB067AB2D514DFF58B189F2CA19},
	doi = {10.1017/S0022377815000495},
	number = {5},
	urldate = {2022-09-19},
	journal = {Journal of Plasma Physics},
	author = {Landreman, Matt and Plunk, Gabriel G. and Dorland, William},
	month = oct,
	year = {2015},
	pages = {905810501}
}

@article{friedman_HW,
	title = {A non-modal analytical method to predict turbulent properties applied to the {Hasegawa}-{Wakatani} model},
	volume = {22},
	issn = {1070-664X},
	url = {https://doi.org/10.1063/1.4905863},
	doi = {10.1063/1.4905863},
	number = {1},
	urldate = {2024-08-16},
	journal = {Physics of Plasmas},
	author = {Friedman, B. and Carter, T. A.},
	month = jan,
	year = {2015},
	pages = {012307}
}

@article{friedman_PRL,
	title = {Linear {Technique} to {Understand} {Non}-{Normal} {Turbulence} {Applied} to a {Magnetized} {Plasma}},
	volume = {113},
	url = {https://link.aps.org/doi/10.1103/PhysRevLett.113.025003},
	doi = {10.1103/PhysRevLett.113.025003},
	number = {2},
	urldate = {2024-08-16},
	journal = {Physical Review Letters},
	author = {Friedman, B. and Carter, T. A.},
	month = jul,
	year = {2014},
	pages = {025003}
}

@article{Arratia2013b,
	author = {Arratia, C. and Chomaz, J. -M.},
	doi = {10.1017/jfm.2013.570},
	title = {On the longitudinal optimal perturbations to inviscid plane shear flow: formal solution and asymptotic approximation},
	journal = {J. Fluid Mech.},
	volume = {737},
	pages = {387-411},
	year = {2013}
}

@article{Butler1992,
	author = {Butler, K. M. and Farrell, B. F.},
	doi = {10.1063.1.858386},
	title = {Three-dimensional optimal perturbations in viscous shear flow},
	journal = {Phys. Fluids A},
	volume = {4},
	number = {8},
	pages = {1637-1650},
	year = {1992}
}

@article{Farrell1988,
	author = {Farrell, B. F.},
	doi = {10.1063/1.866609},
	title = {Optimal excitation of perturbations in viscous shear flow},
	journal = {Phys. Fluids},
	volume = {31},
	number = {8},
	pages = {2093-2102},
	year = {1988}
}

@article{Farrell1993b,
	author = {Farrell, B. F. and Ioannou, P. J.},
	doi = {10.1175/1520-0469(1993)050<2201:TDOPIS>2.0.CO;2},
	title = {Transient development of perturbations in stratified shear flow},
	journal = {J. Atmos. Sci.},
	volume = {50},
	number = {14},
	pages = {2201-2214},
	year = {1993}
}

@phdthesis{Tearle2004,
	author = {Tearle, M. O.},
	title = {Optimal perturbation analysis of stratified shear flows},
	school = {University of Colorado},
	year = {2004}
}

@article{Trefethen1993,
	author = {Trefethen, L. O. and Trefethen, A. E. and Reddy, S. C. and Driscoll, T. A.},
	doi = {10.1126/science.261.5121.578},
	title = {Hydrodynamic stability without eigenvalues},
	journal = {Science},
	volume = {261},
	pages = {578-261},
	year = {1993}
}

@article{pickering_lift-up_2020,
	title = {Lift-up, {Kelvin}–{Helmholtz} and {Orr} mechanisms in turbulent jets},
	volume = {896},
	issn = {0022-1120, 1469-7645},
	url = {https://www.cambridge.org/core/journals/journal-of-fluid-mechanics/article/liftup-kelvinhelmholtz-and-orr-mechanisms-in-turbulent-jets/E02CC1F9FE261D38B2E2A91D77D53E20},
	doi = {10.1017/jfm.2020.301},
	journal = {Journal of Fluid Mechanics},
	author = {Pickering, Ethan and Rigas, Georgios and Nogueira, Petrônio A. S. and Cavalieri, André V. G. and Schmidt, Oliver T. and Colonius, Tim},
	month = aug,
	year = {2020},
	pages = {A2}
}

@article{Terry_ZF,
  title = {Suppression of turbulence and transport by sheared flow},
  author = {Terry, P. W.},
  journal = {Rev. Mod. Phys.},
  volume = {72},
  issue = {1},
  pages = {109--165},
  numpages = {0},
  year = {2000},
  month = {Jan},
  publisher = {American Physical Society},
  doi = {10.1103/RevModPhys.72.109},
  url = {https://link.aps.org/doi/10.1103/RevModPhys.72.109}
}

@article{BNS_KH_PRD,
	title = {Efficient magnetic-field amplification due to the {Kelvin}-{Helmholtz} instability in binary neutron star mergers},
	volume = {92},
	url = {https://link.aps.org/doi/10.1103/PhysRevD.92.124034},
	doi = {10.1103/PhysRevD.92.124034},
	number = {12},
	urldate = {2024-08-15},
	journal = {Physical Review D},
	author = {Kiuchi, Kenta and Cerdá-Durán, Pablo and Kyutoku, Koutarou and Sekiguchi, Yuichiro and Shibata, Masaru},
	month = dec,
	year = {2015},
	pages = {124034}
}

@article{BNS_KH_nature,
	title = {A large-scale magnetic field produced by a solar-like dynamo in binary neutron star mergers},
	volume = {8},
	copyright = {2024 The Author(s)},
	issn = {2397-3366},
	url = {https://www.nature.com/articles/s41550-024-02194-y},
	doi = {10.1038/s41550-024-02194-y},
	number = {3},
	urldate = {2024-08-15},
	journal = {Nature Astronomy},
	author = {Kiuchi, Kenta and Reboul-Salze, Alexis and Shibata, Masaru and Sekiguchi, Yuichiro},
	month = mar,
	year = {2024},
	pages = {298--307}
}

@article{rieger_KH_jet_review,
	title = {An {Introduction} to {Particle} {Acceleration} in {Shearing} {Flows}},
	volume = {7},
	copyright = {http://creativecommons.org/licenses/by/3.0/},
	issn = {2075-4434},
	url = {https://www.mdpi.com/2075-4434/7/3/78},
	doi = {10.3390/galaxies7030078},
	number = {3},
	urldate = {2024-08-15},
	journal = {Galaxies},
	author = {Rieger, Frank M.},
	month = sep,
	year = {2019},
	pages = {78}
}

@article{Perseus_KH,
	title = {Is there a giant {Kelvin}–{Helmholtz} instability in the sloshing cold front of the {Perseus} cluster?},
	volume = {468},
	issn = {0035-8711},
	url = {https://doi.org/10.1093/mnras/stx640},
	doi = {10.1093/mnras/stx640},
	number = {2},
	urldate = {2024-08-15},
	journal = {Monthly Notices of the Royal Astronomical Society},
	author = {Walker, S. A. and Hlavacek-Larrondo, J. and Gendron-Marsolais, M. and Fabian, A. C. and Intema, H. and Sanders, J. S. and Bamford, J. T. and van Weeren, R.},
	month = jun,
	year = {2017},
	pages = {2506--2516}
}

@article{cai_blazar_KH,
	title = {Long-term {Optical} {Monitoring} of the {TeV} {BL} {Lacertae} {Object} {1ES} 2344 + 514},
	volume = {260},
	issn = {0067-0049},
	url = {https://dx.doi.org/10.3847/1538-4365/ac666b},
	doi = {10.3847/1538-4365/ac666b},
	number = {2},
	urldate = {2024-08-15},
	journal = {The Astrophysical Journal Supplement Series},
	author = {Cai, J. T. and Kurtanidze, S. O. and Liu, Y. and Kurtanidze, O. M. and Nikolashvili, M. G. and Xiao, H. B. and Fan, J. H.},
	month = jun,
	year = {2022},
	pages = {47}
}

@article{fan_blazars_KH,
	title = {Characterizing the {Emission} {Region} {Properties} of {Blazars}},
	volume = {268},
	issn = {0067-0049},
	url = {https://dx.doi.org/10.3847/1538-4365/ace7c8},
	doi = {10.3847/1538-4365/ace7c8},
	number = {1},
	urldate = {2024-08-16},
	journal = {The Astrophysical Journal Supplement Series},
	author = {Fan, Junhui and Xiao, Hubing and Yang, Wenxin and Zhang, Lixia and Strigachev, Anton A. and Bachev, Rumen S. and Yang, Jianghe},
	month = sep,
	year = {2023},
	pages = {23}
}

@article{Spiegel_Zahn_Tachocline,
	title = {The solar tachocline.},
	volume = {265},
	issn = {0004-6361},
	url = {https://ui.adsabs.harvard.edu/abs/1992A&A...265..106S},
	urldate = {2024-05-28},
	journal = {Astronomy and Astrophysics},
	author = {Spiegel, E. A. and Zahn, J. -P.},
	month = nov,
	year = {1992},
	pages = {106--114}
}

@article{Garaud_LowPr_review,
	title = {Journey to the center of stars: {The} realm of low {Prandtl} number fluid dynamics},
	volume = {6},
	shorttitle = {Journey to the center of stars},
	url = {https://link.aps.org/doi/10.1103/PhysRevFluids.6.030501},
	doi = {10.1103/PhysRevFluids.6.030501},
	number = {3},
	urldate = {2023-02-06},
	journal = {Physical Review Fluids},
	author = {Garaud, P.},
	month = mar,
	year = {2021},
	pages = {030501}
}

@article{burrell_effects_1997,
	title = {Effects of {E}×{B} velocity shear and magnetic shear on turbulence and transport in magnetic confinement devices},
	volume = {4},
	issn = {1070-664X},
	url = {https://doi.org/10.1063/1.872367},
	doi = {10.1063/1.872367},
	number = {5},
	urldate = {2024-08-19},
	journal = {Physics of Plasmas},
	author = {Burrell, K. H.},
	month = may,
	year = {1997},
	pages = {1499--1518}
}

@book{zhou_hydrodynamic_2024,
	address = {Cambridge, United Kingdom ; New York, NY},
	edition = {1st edition},
	title = {Hydrodynamic {Instabilities} and {Turbulence}: {Rayleigh}–{Taylor}, {Richtmyer}–{Meshkov}, and {Kelvin}–{Helmholtz} {Mixing}},
	isbn = {978-1-108-48964-5},
	shorttitle = {Hydrodynamic {Instabilities} and {Turbulence}},
	publisher = {Cambridge University Press},
	author = {Zhou, Ye},
	month = dec,
	year = {2024},
}

@article{zhou_instabilities_2024,
	title = {Instabilities and {Mixing} in {Inertial} {Confinement} {Fusion}},
	url = {https://www.annualreviews.org/content/journals/10.1146/annurev-fluid-022824-110008},
        journal = {Annual Reviews of Fluid Mechanics},
	doi = {10.1146/annurev-fluid-022824-110008},
	urldate = {2025-01-14},
	author = {Zhou, Ye and Sadler, James D. and Hurricane, Omar A.},
	month = sep,
	year = {2024}
}

@article{zhou_rayleightaylor_2017,
	series = {Rayleigh–{Taylor} and {Richtmyer}–{Meshkov} instability induced flow, turbulence, and mixing. {II}},
	title = {Rayleigh–{Taylor} and {Richtmyer}–{Meshkov} instability induced flow, turbulence, and mixing. {II}},
	volume = {723-725},
	issn = {0370-1573},
	url = {https://www.sciencedirect.com/science/article/pii/S0370157317302958},
	doi = {10.1016/j.physrep.2017.07.008},
	urldate = {2025-01-14},
	journal = {Physics Reports},
	author = {Zhou, Ye},
	month = dec,
	year = {2017},
	pages = {1--160}
}

@article{chow_kelvinhelmholtz_2023,
	title = {The {Kelvin}–{Helmholtz} {Instability} at the {Boundary} of {Relativistic} {Magnetized} {Jets}},
	volume = {951},
	issn = {2041-8205},
	url = {https://dx.doi.org/10.3847/2041-8213/acdfcf},
	doi = {10.3847/2041-8213/acdfcf},
	number = {2},
	urldate = {2024-08-27},
	journal = {The Astrophysical Journal Letters},
	author = {Chow, Anthony and Davelaar, Jordy and Rowan, Michael E. and Sironi, Lorenzo},
	month = jul,
	year = {2023},
	pages = {L23}
}

@article{Ref3_1,
	title = {Amplification of small perturbations in a {Hartmann} layer},
	volume = {14},
	issn = {1070-6631},
	url = {https://doi.org/10.1063/1.1456512},
	doi = {10.1063/1.1456512},
	number = {4},
	urldate = {2025-09-11},
	journal = {Physics of Fluids},
	author = {Gerard-Varet, D.},
	month = apr,
	year = {2002},
	pages = {1458--1467},
}

@article{Ref3_2,
	title = {Quasi-two-dimensional perturbations in duct flows under transverse magnetic field},
	volume = {19},
	issn = {1070-6631},
	url = {https://doi.org/10.1063/1.2747233},
	doi = {10.1063/1.2747233},
	number = {7},
	urldate = {2025-09-11},
	journal = {Physics of Fluids},
	author = {Pothérat, A.},
	month = jul,
	year = {2007},
	pages = {074104},
}

@article{Ref3_3,
	title = {Optimal linear growth in magnetohydrodynamic duct flow},
	volume = {653},
	issn = {1469-7645, 0022-1120},
	url = {https://www.cambridge.org/core/journals/journal-of-fluid-mechanics/article/abs/optimal-linear-growth-in-magnetohydrodynamic-duct-flow/2278003F4AE59B160BCB735755663954},
	doi = {10.1017/S0022112010000273},
	urldate = {2025-09-11},
	journal = {Journal of Fluid Mechanics},
	author = {Krasnov, Dmitry and Zikanov, Oleg and Rossi, Maurice and Boeck, Thomas},
	month = jun,
	year = {2010},
	pages = {273--299},
}

@article{Ref3_4,
	title = {Laminar-{Turbulent} {Transition} in {Magnetohydrodynamic} {Duct}, {Pipe}, and {Channel} {Flows}},
	volume = {66},
	issn = {0003-6900},
	url = {https://doi.org/10.1115/1.4027198},
	doi = {10.1115/1.4027198},
	number = {030802},
	urldate = {2025-09-11},
	journal = {Applied Mechanics Reviews},
	author = {Zikanov, Oleg and Krasnov, Dmitry and Boeck, Thomas and Thess, Andre and Rossi, Maurice},
	month = apr,
	year = {2014},
}

@article{Ref3_5,
	title = {Optimal transient growth and transition to turbulence in the {MHD} pipe flow subject to a transverse magnetic field},
	volume = {9},
	doi = {10.1103/PhysRevFluids.9.103702},
	number = {10},
	journal = {Physical Review Fluids},
	author = {Velizhanina, Yelyzaveta},
	year = {2024},
}

@article{Ref3_6,
	title = {Optimal growth and transition to turbulence in channel flow with spanwise magnetic field},
	volume = {596},
	issn = {1469-7645, 0022-1120},
	url = {https://www.cambridge.org/core/journals/journal-of-fluid-mechanics/article/abs/optimal-growth-and-transition-to-turbulence-in-channel-flow-with-spanwise-magnetic-field/0D10CCEADBCD5A90F57ADE1C9F3B211F},
	doi = {10.1017/S002211200700924X},
	urldate = {2025-09-11},
	journal = {Journal of Fluid Mechanics},
	author = {Krasnov, Dmitry and Rossi, Maurice and Zikanov, Oleg and Boeck, Thomas},
	month = jan,
	year = {2008},
	pages = {73--101},
}

@article{Ref3_7,
	title = {Secondary energy growth and turbulence suppression in conducting channel flow with streamwise magnetic field},
	volume = {24},
	issn = {1070-6631},
	url = {https://doi.org/10.1063/1.4731293},
	doi = {10.1063/1.4731293},
	number = {7},
	urldate = {2025-09-11},
	journal = {Physics of Fluids},
	author = {Dong, Shuai and Krasnov, Dmitry and Boeck, Thomas},
	month = jul,
	year = {2012},
	pages = {074101},
}

@article{Ref3_8,
	title = {Optimal perturbations and transition in the magnetohydrodynamic boundary layer under the influence of a spanwise magnetic field},
	volume = {34},
	issn = {1070-6631},
	url = {https://doi.org/10.1063/5.0089403},
	doi = {10.1063/5.0089403},
	number = {5},
	urldate = {2025-09-11},
	journal = {Physics of Fluids},
	author = {Bourcy, S. and Velizhanina, Y. and Pavlenko, Y. and Knaepen, B.},
	month = may,
	year = {2022},
	pages = {054115},
}

@article{Ref3_9,
	title = {From three-dimensional to quasi-two-dimensional: transient growth in magnetohydrodynamic duct flows},
	volume = {861},
	issn = {0022-1120, 1469-7645},
	shorttitle = {From three-dimensional to quasi-two-dimensional},
	url = {https://www.cambridge.org/core/journals/journal-of-fluid-mechanics/article/abs/from-threedimensional-to-quasitwodimensional-transient-growth-in-magnetohydrodynamic-duct-flows/0E1AE14E3CEA9830F047BD4DA306F0B9},
	doi = {10.1017/jfm.2018.863},
	urldate = {2025-09-11},
	journal = {Journal of Fluid Mechanics},
	author = {Cassells, Oliver G. W. and Vo, Tony and Pothérat, Alban and Sheard, Gregory J.},
	month = feb,
	year = {2019},
	pages = {382--406}
}

@article{horn_2022,
	title = {The {Elbert} range of magnetostrophic convection. {I}. {Linear} theory},
	volume = {478},
	issn = {1364-5021},
	url = {https://doi.org/10.1098/rspa.2022.0313},
	doi = {10.1098/rspa.2022.0313},
	number = {2264},
	urldate = {2025-12-09},
	journal = {Proceedings of the Royal Society A: Mathematical, Physical and Engineering Sciences},
	author = {Horn, Susanne and Aurnou, Jonathan M.},
	month = aug,
	year = {2022},
	pages = {20220313}
}

@article{lalloz_alfven_2025,
	title = {Alfvén waves at low magnetic {Reynolds} number: transitions between diffusion, dispersive {Alfvén} waves and nonlinear propagation},
	volume = {1003},
	issn = {0022-1120, 1469-7645},
	shorttitle = {Alfvén waves at low magnetic {Reynolds} number},
	url = {https://www.cambridge.org/core/journals/journal-of-fluid-mechanics/article/alfven-waves-at-low-magnetic-reynolds-number-transitions-between-diffusion-dispersive-alfven-waves-and-nonlinear-propagation/9E643A0C7B463003F86FDDE448561165},
	doi = {10.1017/jfm.2024.1165},
	urldate = {2025-12-10},
	journal = {Journal of Fluid Mechanics},
	author = {Lalloz, Samy and Davoust, Laurent and Debray, François and Pothérat, Alban},
	month = jan,
	year = {2025},
	pages = {A19}
}

@ARTICLE{Skoutnev,
       author = {{Skoutnev}, Valentin A. and {Beloborodov}, Andrei M.},
        title = "{Tayler Instability Revisited}",
      journal = {The Astrophysical Journal},
         year = 2024,
        month = oct,
       volume = {974},
       number = {2},
          eid = {290},
        pages = {290},
          doi = {10.3847/1538-4357/ad71c8},
archivePrefix = {arXiv},
       eprint = {2404.19103},
 primaryClass = {astro-ph.SR},
       adsurl = {https://ui.adsabs.harvard.edu/abs/2024ApJ...974..290S}
}

\end{document}